# Large perpendicular magnetic anisotropy of transition metal dimers driven by polarization switching of two-dimensional ferroelectric In$_2$Se$_3$ substrate

Wen Qiao,[1] Deyou Jin,[1] Wenbo Mi,[1,2] Dunhui Wang,[1] Shiming Yan,[1,*] Xiaoyong Xu,[1,3,*] Tiejun Zhou[1,*]

[1] *School of Electronics and Information, Hangzhou Dianzi University, Hangzhou 310018, China*

[2] *Tianjin Key Laboratory of Low Dimensional Materials Physics and Preparation Technology, School of Science, Tianjin University, Tianjin 300354, China*

[3] *School of Physics Science and Technology, Yangzhou University, Yangzhou, 225002, China*

[*]Author to whom all correspondence should be addressed.
E-mail: shimingyan@hdu.edu.cn, xxy@yzu.edu.cn, tjzhou@hdu.edu.cn

# ABSTRCT

Large perpendicular magnetic anisotropy (MA) is highly desirable for realizing atomic-scale magnetic data storage which represents the ultimate limit of the density of magnetic recording. In this work, we studied the MA of transition metal dimers Co-Os, Co-Co and Os-Os adsorbed on two-dimensional ferroelectric $In_2Se_3$ ($In_2Se_3$-CoOs, $In_2Se_3$-OsCo, $In_2Se_3$-CoCo and $In_2Se_3$-OsOs) by first-principles calculations. It is found that the Co-Os dimer in $In_2Se_3$-CoOs has large total perpendicular magnetic anisotropy energy (MAE) of ~ 40 meV. In particular, the MAE arising from Os atom is up to ~ 60 meV. The large MAE is attributed to the high spin-orbit coupling constant and the onefold coordination of Os atom. In addition, the MA of the dimers can be tuned by the polarization reversal of $In_2Se_3$. When the polarization is upward (P↑), the easy-axis directions of MA in $In_2Se_3$-OsCo, $In_2Se_3$-CoCo and $In_2Se_3$-OsOs are all in-plane, while the directions become perpendicular as the polarization is switched to downward (P↓). For the $In_2Se_3$-CoOs, switching polarization from P↑ to P↓ enhance the perpendicular MA from ~ 20 meV to ~ 40 meV. Based on the second-order perturbation theory, we confirm that the exchange splitting of $d_{xy}/d_{x2-y2}$ and $d_{xz}/d_{yz}$ orbitals as well as the occupation of $d_{z2}$ orbital at the vicinity of Fermi level play important roles in the changes of MA with the reversal of FE polarization of $In_2Se_3$.

## 1. INTRODUCTION

With the rapid development of electronic industry ultrahigh-density data storage technology is attracting more and more attention from researchers. For magnetic data storage, a direct way to improve the density of data is to reduce the size of recorded bit unit, which will reach the theoretical limit when the size decreases to atomic scale.[1-5] One of the key requirements for achieving the atomic-scale information storage is to endow the atomic data bit with a large magnetic anisotropy (MA).[1,2] In particular, a large perpendicular MA is more desirable because of its uniaxial feature. The larger the magnetic anisotropy energy (MAE), the more stable the stored information is. However, when the information bits of the magnetic atoms are wrote or manipulated, the large MA often requires a strong magnetic field or a high flip current, which is detrimental to the low power consumption and consequently not facilitate to the miniaturization of the data-storage device.[6,7] If the MA can be tuned in situ, the energy consumption required for information writing and manipulating can be reduced. Therefore, exploring large and tunable MA is significant for the applications of atom-scale magnets in magnetic storage and logical devices.

Large MA usually stems from spin-orbit coupling (SOC).[8,9] Orbital momentum is indispensable in inducing the MA. Free transition metal atoms have high orbital momentum, but the energies of states with different directions of orbital momentum are identical. There is no MA present in a freestanding atom. By means of atom coordination the orbital momentums of partial directions can be quenched or reduced, consequently inducing MA. In solids, however, the coordination of an inner atom is highly symmetrical, which results in low MA. High MA usually appears in a low symmetry or low coordination system. It has been found that the transition metal dimers, the smallest size of the transition metal clusters, such as Rh-Rh, Ir-Ir and Pt-Pt show high MAEs.[10-13] However, the freestanding dimers possess no functionality in electric device. For practical application, a suitable substrate should be required to support the dimers. In addition, to tune the MA, it is also more convenient to realize through the substrate.

Single-layer α-In$_2$Se$_3$ is a 2D ferroelectric material, which is first predicted by first-principles density-function theory (DFT) calculations and subsequently be successfully prepared and confirmed in experiment.[14-17] Due to the unique geometric structure and large surface area, monolayer α-In$_2$Se$_3$ provides a suitable platform for supporting nanostructures. Furthermore, as the asymmetric Se atoms in the middle of the layer can spontaneously break the centrosymmetry, monolayer α-In$_2$Se$_3$ show highly stable out-of-plane electric polarization at room temperature. The highest potential barrier is about 66 meV per unit lattice.[15] This offers a possibility to tune the magnetism of the supported nanostructures via the ferroelectric (FE) polarization of the α-In$_2$Se$_3$ substrate. Recently, in 2D FeI$_2$/In$_2$Se$_3$ heterostructures, it is found that the FE polarization of In$_2$Se$_3$ can induce a transition of FeI$_2$ from ferromagnetic coupling to antiferromagnetic coupling.[18] In 2D In$_2$Se$_3$/Cr$_2$Ge$_2$Te$_6$ heterostructures, the MA of Cr$_2$Ge$_2$Te$_6$ can vary between the out-of-plane and in-plane orientations with the changes of the polarization direction of In$_2$Se$_3$.[19] In addition to the 2D heterostructures, strong magnetoelectric effects were also found in the systems of single molecule magnets adsorbed on the In$_2$Se$_3$ substrate. The magnetism, including magnetic moments and MA, in metal phthalocyanine molecules can be efficiently manipulated by the FE polarization of In$_2$Se$_3$.[20,21] Similarly, in metal porphyrazine molecules, it is also found that the FE polarization of In$_2$Se$_3$ can effectively tune the magnetic moments.[22] For magnetic dimers, thus far, there is no report on tuning of magnetism via the FE polarization of In$_2$Se$_3$.

In this work, we studied the MA of Co-Os dimer on 2D In$_2$Se$_3$. Co atom has high intra-atom exchange interaction but low SOC constant; while Os atom has large SOC constant, but its intra-atom exchange interaction tends to be reduced by orbital hybridization with nonmagnetic atoms. Combination of these two transition metal atoms is expected to give rise to high SOC, and hence probably inducing large MA. For comparison, we also studied the dimers, Co-Co and Os-Os. The calculated results show that the Co-Os dimer on In$_2$Se$_3$ substrate indeed shows high total MAE, ~ 40 meV. The MAE arising from Os atom is up to ~ 60 meV. Such large MAE is attributed to the high SOC constant and the onefold coordination of Os atom. We further studied the influence of the polarization reversal of In$_2$Se$_3$ on the magnetic moments and MA for the adsorbed dimers. Based on the second-order perturbation theory, we analysed the roles exchange splitting and distribution of d orbitals around Fermi level in the changes of MAE with reversal of FE polarization of In$_2$Se$_3$.

## 2. METHODS OF CALCULATION

All calculations are performed within the framework of DFT implemented in the Vienna ab initio simulation package with the Perdew-Burke-Ernzerhor functional in the generalized gradient approximation. The ion-electron interaction was described by the projector-augmented plane-wave potentials. A plane wave basis set with a kinetic energy cutoff of 400 eV was employed. We used a 2×2 In$_2$Se$_3$ super cell with a vacuum slab 20 Å for minimizing image interaction from periodic boundary condition. The positions of all atoms were relaxed until the force was less than 0.01 eV/Å. The convergence criterion for the total energy was set to 1×10$^{-6}$ eV. The binding energy was calculated using the following formula, $E_b = E_{total} - E_{monolayer} - E_{atom}$. The SOC included in the noncollinear calculations using the second variation method is considered for calculating the MAE.[23,24] The MAE is obtained based on the total energy difference when the magnetization directions are in the $xy$ plane ($E^{\parallel}$) and along the $z$ axis ($E^{\perp}$), MAE = $E^{\parallel} - E^{\perp}$. Positive and negative values of MAE show the easy magnetization axis along the out-of-plane and in-plane directions, respectively.

## 3. RESULTS AND DISCUSSION

On the surface of α-In$_2$Se$_3$, there are three different sites, hole site, In atop site and Se atop site, as shown in Figure 1(a). We first explore the most stable adsorption site for the transition metal dimers. We considered four systems with dimers vertically adsorbed on In$_2$Se$_3$, In$_2$Se$_3$-CoCo, In$_2$Se$_3$-OsOs, In$_2$Se$_3$-CoOs and In$_2$Se$_3$-OsCo [Figure 1(b)]. For In$_2$Se$_3$-CoCo, the upper Co atom in dimer is labeled as Co2; the lower Co atom that is contiguous to the basal plane of In$_2$Se$_3$ is labeled as Co1, as shown in Figure 1(b). The numbers in Os1 and Os2 of In$_2$Se$_3$-OsOs represent the same meaning as Co1 and Co2 in In$_2$Se$_3$-CoCo. With respect to the Co-Os dimer, there are two configurations, In$_2$Se$_3$-CoOs and In$_2$Se$_3$-OsCo. In the two configurations, the Os atoms are situated at the top site (In$_2$Se$_3$-CoOs) and bottom site (In$_2$Se$_3$-OsCo), respectively. The four systems with the ferroelectric polarization upward (P↑) and downward (P↓) were optimized and the corresponding binding energies of the dimers were calculated. In addition to the vertical configurations, we also tried to set the dimers parallel to the basal plane of In$_2$Se$_3$. However, after optimization the parallel dimers separated into two individual atoms at different sites on In$_2$Se$_3$. Here, we studied the vertical configuration of dimers, since after optimization the two atoms of the dimers with this configuration are still combined together. The calculated binding energies for the dimers with vertical configurations are shown in Table 1. These values indicate that, for both P↑ and P↓ of In$_2$Se$_3$-CoCo, In$_2$Se$_3$-OsOs, In$_2$Se$_3$-CoOs and In$_2$Se$_3$-OsCo, the hole site is the most stable position of the dimers relative to the other two sites, implying that under reversing of FE polarization the dimers will be fixed in the hole sites.

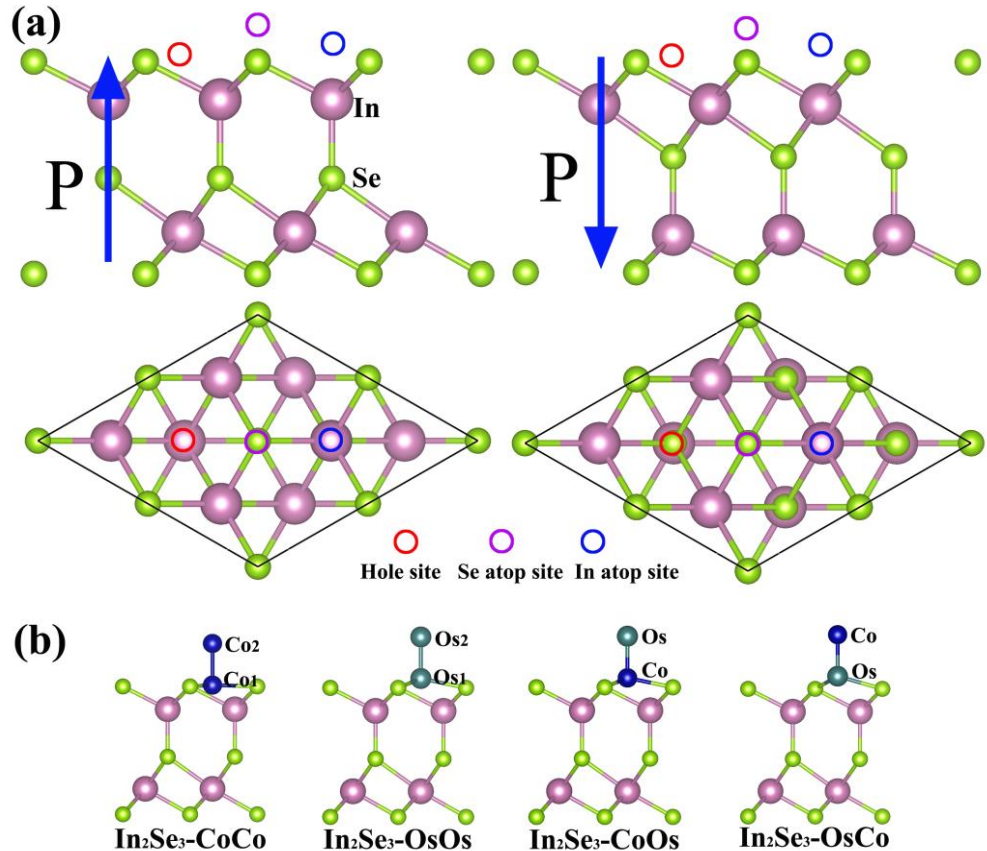

**Figure 1** (a) Top and side views of In$_2$Se$_3$ supercell with FE polarization P↑ and P↓. (b) Optimized configurations of dimers, Co-Co, Os-Os, Co-Os, vertically adsorbed on In$_2$Se$_3$.

Table 1. The binding energies of the atoms (Co and Os) and the dimers (Co-Co, Os-Os, Co-Os) at the hole site, In atop site and Se atop site of $In_2Se_3$ with FE polarization P↑ and P↓.

|  | Hole site (eV) | | In atop site (eV) | | Se atop site (eV) | |
| --- | --- | --- | --- | --- | --- | --- |
|  | P↑ | P↓ | P↑ | P↓ | P↑ | P↓ |
| **Co** | -1.82 | -2.07 | -0.25 | -0.98 | 0.02 | 0.65 |
| **Os** | -1.85 | -1.95 | -0.32 | -1.35 | 0.04 | 0.45 |
| **Co-Co** | -1.96 | -2.23 | -0.61 | -0.87 | -0.35 | -0.23 |
| **Os-Os** | -0.48 | -0.86 | -0.07 | -0.57 | -0.13 | -0.08 |
| **Co-Os** | -2.21 | -2.58 | -1.63 | -2.27 | -1.37 | -1.01 |
| **Os-Co** | -2.95 | -3.33 | -2.05 | -2.93 | -1.85 | -1.49 |

Based on the calculated results of the binding energies, we chose the configurations with dimers adsorbed on the hole sites to further explore the magnetism and the effects of FE polarization. The calculated magnetic moments of the dimers with P↑ and P↓ of $In_2Se_3$ are shown in Figure 2(a). It can be seen that the magnetic moments of the upper atoms (Co2 in $In_2Se_3$-CoCo, Os2 in $In_2Se_3$-OsOs, Os in $In_2Se_3$-CoOs, and Co in $In_2Se_3$-OsCo) in dimers are larger than those of the lower atoms (Co1 in $In_2Se_3$-CoCo, Os1 in $In_2Se_3$-OsOs, Co in $In_2Se_3$-CoOs, and Os in $In_2Se_3$-OsCo). This can be attributed to the lower coordination for the upper atoms. The coordination number of the lower atoms is 4, belonging to fourfold coordination; while the coordination number of the upper atom is 1, belonging to onefold coordination. Coordination will induce ligand field effects and electron delocalization which will broaden energy bands and reduce the magnetic moments.[2] Therefore, the upper atoms in dimers possess larger magnetic moments. The FE polarization has slightly influence on the magnetic moments of the dimers. For the lower atoms, the magnetic moments of atoms in the four systems with P↓ are all larger than those with P↑. For the upper atoms, the magnetic moments of Os2 in $In_2Se_3$-OsOs and Co in $In_2Se_3$-OsCo follow the same law as the lower atoms. Whereas, for the upper atoms Co2 in $In_2Se_3$-CoCo and Os in $In_2Se_3$-CoOs, the magnetic moments reduce as the FE polarization changes from P↑ to P↓ [see Figure 2(a)].

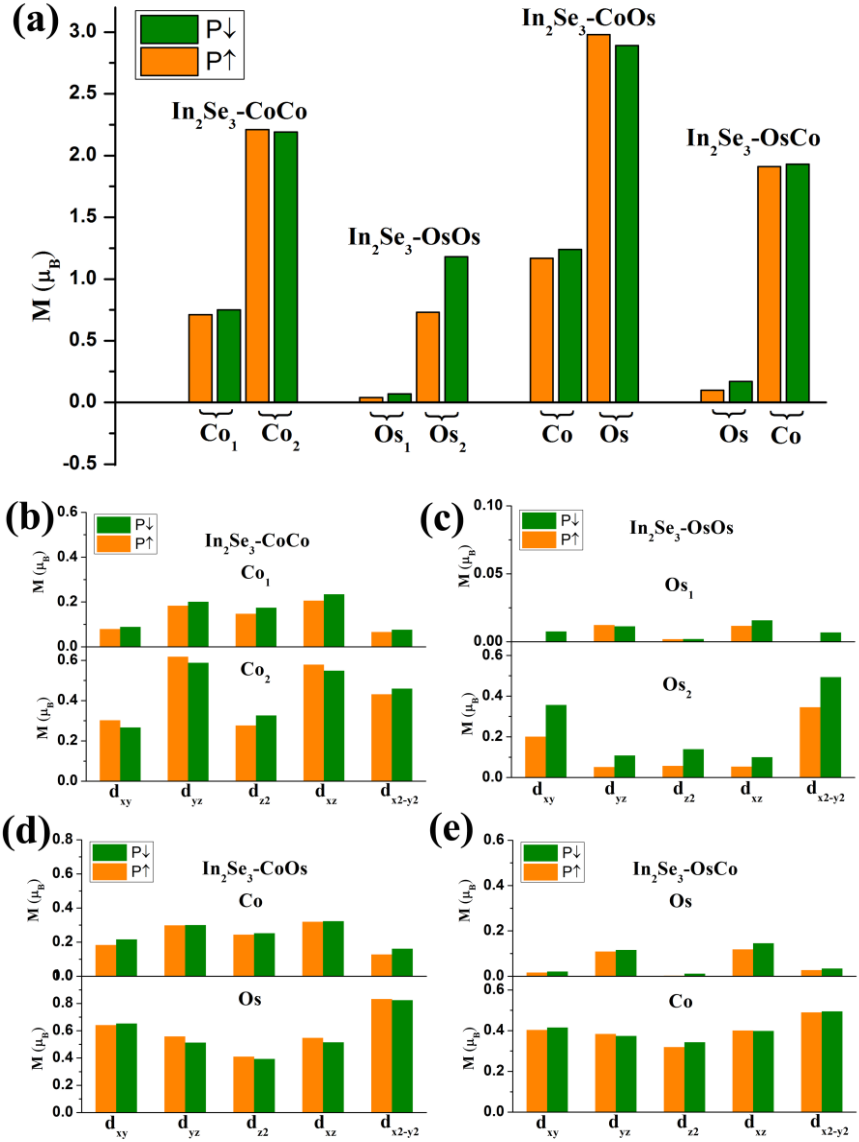

**Figure 2** (a) Magnetic moments of atoms in dimers of In$_2$Se$_3$-CoCo, In$_2$Se$_3$-OsOs, In$_2$Se$_3$-CoOs, and In$_2$Se$_3$-OsCo with FE polarization P↑ and P↓. Orbital-resolved magnetic moments of (b) In$_2$Se$_3$-CoCo, (c) In$_2$Se$_3$-OsOs, (d) In$_2$Se$_3$-CoOs, and (e) In$_2$Se$_3$-OsCo with FE polarization P↑ and P↓.

To get insight of the origin of magnetic moments, we calculated the orbital-resolved density of states (DOS). From the DOS, we further calculated the orbital-resolved magnetic moments. As shown in Figures 2(b)-(e), it can be seen that the $d_{xz}$ and $d_{yz}$ orbitals contribute the main magnetic moments for the lower atoms in the four systems (In$_2$Se$_3$-CoCo, In$_2$Se$_3$-OsOs, In$_2$Se$_3$-CoOs and In$_2$Se$_3$-OsCo). The magnetic moments of the upper atom Co2 in In$_2$Se$_3$-CoCo also mainly come from the

$d_{xz}$ and $d_{yz}$ orbitals; while the magnetic moments of upper atoms in other three systems mainly stem from the $d_{xy}$ and $d_{x2-y2}$ orbitals. For both upper and lower atoms, the $d_{z2}$ orbitals all contribute the smallest value for total magnetic moments in the four systems. As shown in Figure 3, the states of $d_{z2}$ orbitals occupy the lowest energy level. This leads to preferentially filling of electrons in this orbital of spin-minority bands, resulting in the low contribution to magnetic moments. FE polarization has almost the same effects on the magnetic moments for all the d orbitals of the lower atoms in the four systems. When the FE polarization switches from P↑ to P↓, the magnetic moments from d orbitals in the lower atoms all become larger. Similar to lower atoms, the upper atom Os2 in In$_2$Se$_3$-OsOs also show significantly enhancement with reversal of FE polarization from P↑ to P↓. For other upper atoms, the FE polarization has different influence on the magnetic moments for different d orbitals. It is noted that, in contrast to the lower atoms, the magnetic moments of d$_{xz}$ and d$_{yz}$ orbitals of the upper atoms in In$_2$Se$_3$-CoCo, In$_2$Se$_3$-CoOs, and In$_2$Se$_3$-OsCo show opposite changes with FE polarization reversing. The changes of exchange splitting are responsible for the dependence of magnetic moments on FE polarization. From the Figure 3(b), it can be clearly seen that, with FE polarization changing from P↑ to P↓, the exchange splitting of Os2 atom is enlarged, which gives rise to the larger magnetic moment.

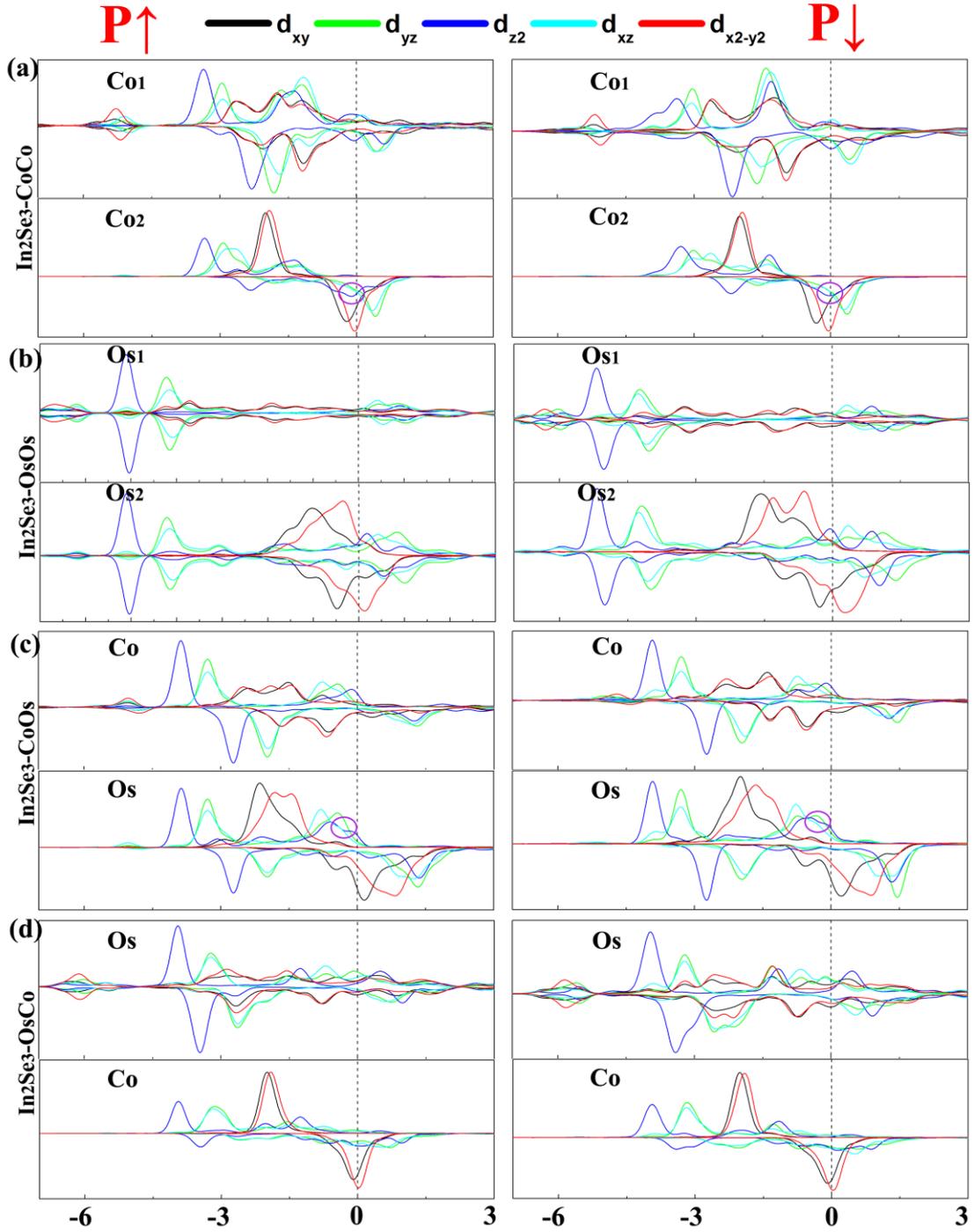

**Figure 3** Orbital-resolved density of states of (a) $In_2Se_3$-CoCo, (b) $In_2Se_3$-OsOs, (c) $In_2Se_3$-CoOs and (d) $In_2Se_3$-OsCo with FE polarization P↑ and P↓.

The total MAEs for the four systems with P↑ and P↓ are shown in Figure 4(a). The four systems all show increase of MAE with the FE polarization switching from P↑ to P↓. The In$_2$Se$_3$-CoOs with P↓ has the largest value (~ 40 meV) of MAE, which is larger than thermal excitation energy of 26 meV (obtained by $K_b$ T, $K_b$ is Boltzmann constant, T = 300 K) at room temperature. The MAE of In$_2$Se$_3$-CoCo with P↓ is ~ 10 meV. As shown in Figure 4(a), this value is lower than those of dimers, Os-Os and Co-Os, which contain Os atom, indicating that combining Os atom can enhance the MAE due to its large SOC constant. It is further noted that the MAE of In$_2$Se$_3$-CoCo is larger than that (~ 1.5 meV) of single Co adsorbed on In$_2$Se$_3$ (In$_2$Se$_3$-Co), confirming the advantage of magnetic dimers in inducing large MA. As shown in Figure 4(a), the MA including not only the strength but also the direction can be profoundly regulated by the FE polarization. For In$_2$Se$_3$-CoCo, In$_2$Se$_3$-OsOs, and In$_2$Se$_3$-OsCo, the directions of MA change from in-plane to perpendicularity with reversal of FE polarization from P↑ to P↓. For the system In$_2$Se$_3$-CoOs, the total MA with P↑ and P↓ are both positive, showing perpendicular easy axis. When the FE polarization changes from P↑ to P↓, the perpendicular MAE becomes larger.

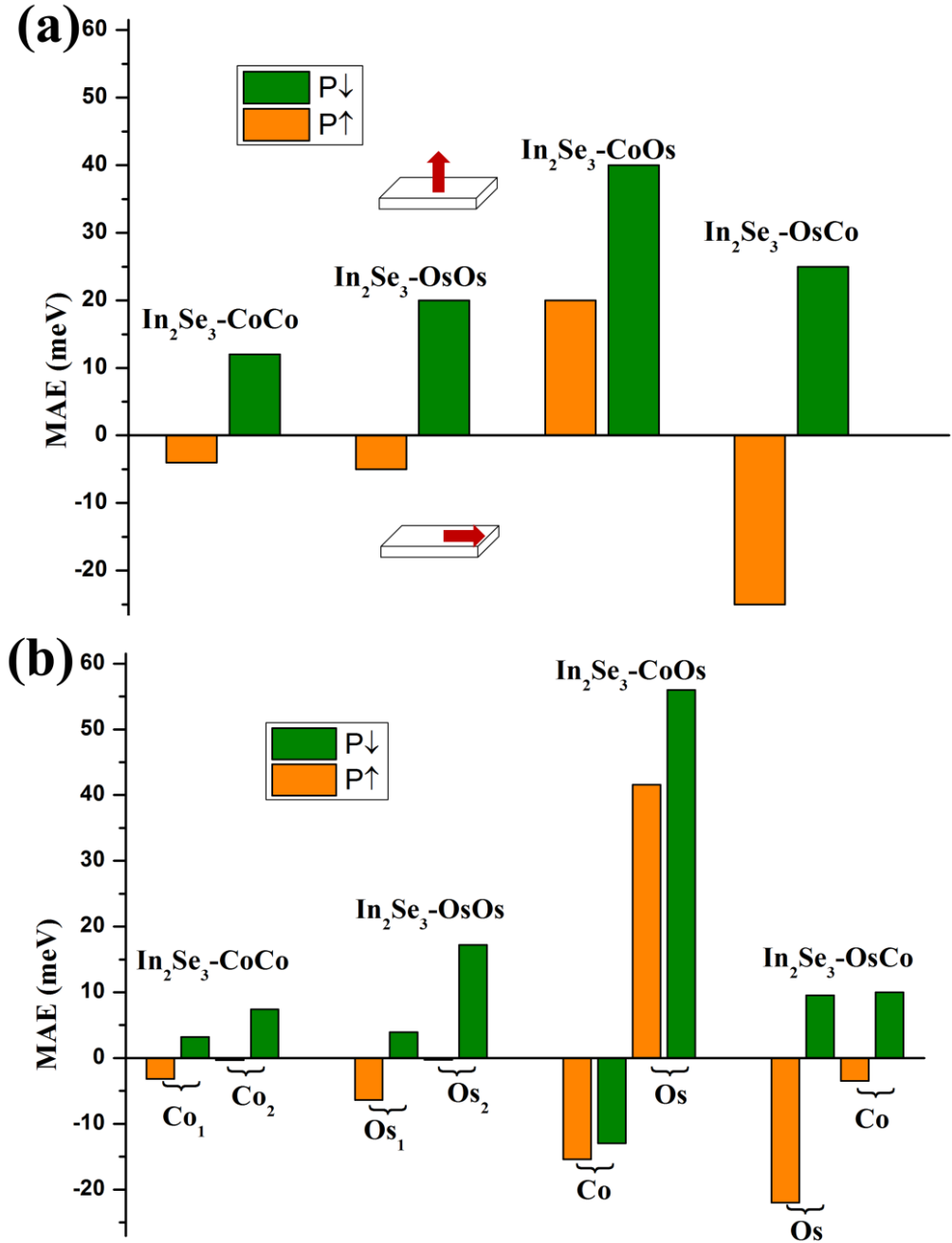

**Figure 4** (a) Total MAE and (b) atom-resolved MAE of In$_2$Se$_3$-CoCo, In$_2$Se$_3$-OsOs, In$_2$Se$_3$-CoOs and In$_2$Se$_3$-OsCo with FE polarization P↑ and P↓.

Figure 4(b) shows the atom-resolved MAE of the four systems with P↑ and P↓. For the system In$_2$Se$_3$-CoOs, the Co atom contributes negative MAE; while the Os atom contributes the positive MAE. Because the absolute value of positive MAE from Os is larger than that of the negative MAE from Co, the total MAE is positive and suggests the out-of-plane easy axis. In contrast to the P↑, P↓ reduces the in-plane MAE of Co atom and enhances the out-of-plane MAE of Os atoms, resulting in the enhancement of the total MAE. Significantly, the MAE for Os with P↓ in In$_2$Se$_3$-CoOs is up to ~ 60 meV. Such large MAE results from the large SOC constant of Os and the special configuration with the onefold coordination. In In$_2$Se$_3$-CoCo and In$_2$Se$_3$-OsOs with P↓, the upper atoms both show larger MAE compared with the lower atom, confirming that the onefold coordination plays an important role in the large MAE.

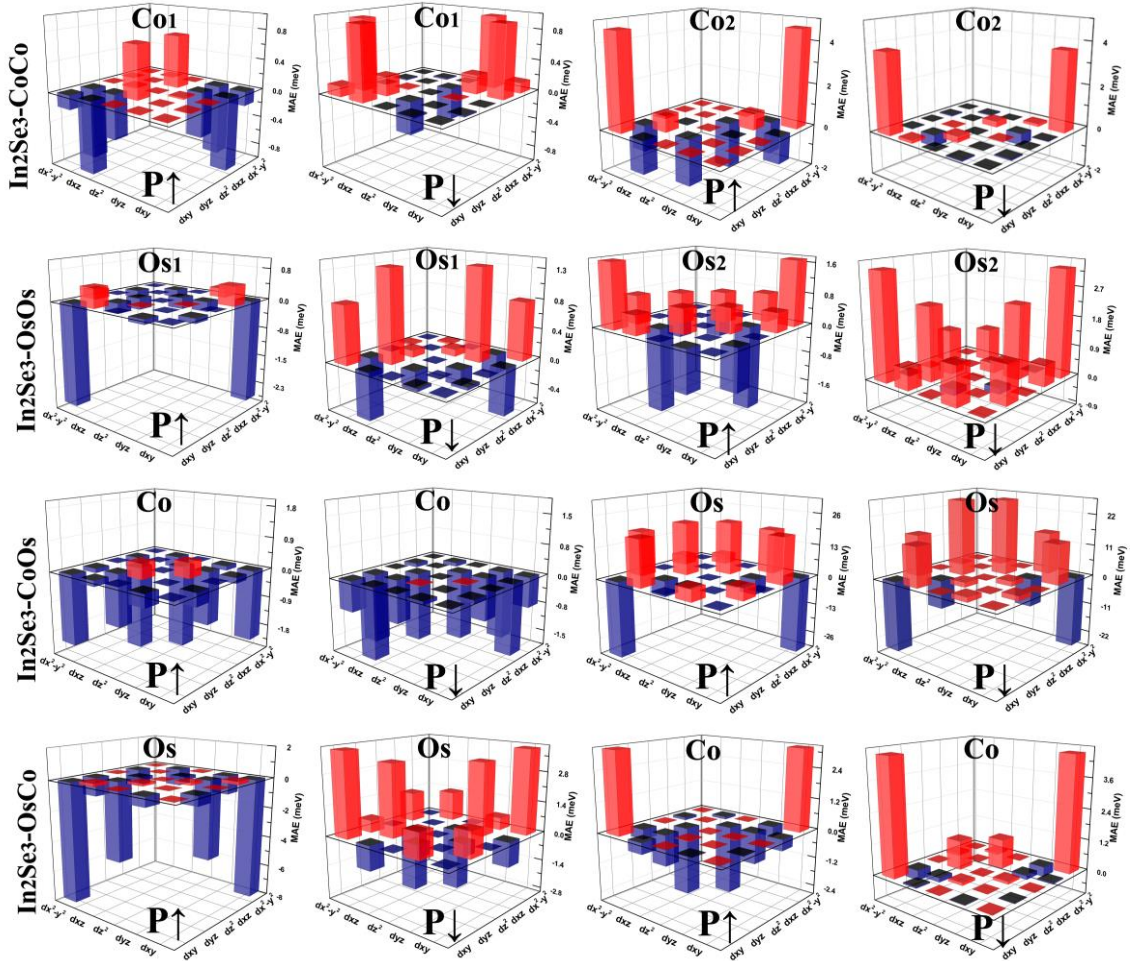

**Figure 5** Orbital-resolved MAE of In$_2$Se$_3$-CoCo, In$_2$Se$_3$-OsOs, In$_2$Se$_3$-CoOs and In$_2$Se$_3$-OsCo with FE polarization P↑ and P↓.

We further explore the orbital-resolved MA in these systems. We calculated the orbital-resolved MAE and analysed the contribution of orbital to the MA. As shown in Figure 5, the diagonal values of the MAE matrix are zero, indicating that the individual d orbitals have no MA contribution. This is because the d orbitals, $d_{xy}$, $d_{xz}$, $d_{z2}$, $d_{yz}$, and $d_{x2-y2}$ are real orbitals. There are no orbital momentums on these single orbitals. The MA is only formed by the real orbital mixture under SOC. From the Figure 5, we can see that, except for Co1 atom, the mixed in-plane obtials $d_{xy}/d_{x2-y2}$ of the other atoms in the four systems all have relatively high proportion in contribution to the total MAE. The small MAE of $d_{xy}/d_{x2-y2}$ orbital in Co1 atoms can be attributed to the low distribution of $d_{xy}/d_{x2-y2}$ state at the vicinity of Fermi level [see Figure 3(a)]. The changes of MAEs from $d_{xy}/d_{x2-y2}$ and $d_{xz}/d_{yz}$ orbitals with FE polarization are responsible for the changes of total MAEs for Os1 and Os2 atoms in $In_2Se_3$-OsOs system, Co atom in $In_2Se_3$-CoOs system, as well as Os and Co atoms in $In_2Se_3$-OsCo system. For the $In_2Se_3$-CoCo system, the change of MA direction for Co1 atom mainly arise from the mixed orbitals, $d_{xz}/d_{xy}$ and $d_{x2-y2}/d_{yz}$; for the Co2 atom, the increase of MAE from $d_{z2}/d_{yz}$ and $d_{z2}/d_{xz}$ orbitals dominate the change of MA direction. In respect to the Os atom in $In_2Se_3$-CoOs system, the alteration of MA direction mainly stems from the $d_{xy}/d_{x2-y2}$ and $d_{z2}/d_{xz}$ orbitals.

In order to deeply understand the influence of FE polarization reversal on the MAE of the dimers on $In_2Se_3$, we analysed the MAE based on the second-order perturbation theory, the MAE caused by SOC is expressed by [25]:

$$\Delta E^{\downarrow\downarrow} = E^{\downarrow\downarrow}(x) - E^{\downarrow\downarrow}(z)$$

$$= \xi^2 \sum_{o^\downarrow, u^\downarrow} \frac{\left|\langle o^\downarrow |L_z| u^\downarrow \rangle\right|^2 - \left|\langle o^\downarrow |L_x| u^\downarrow \rangle\right|^2}{E_u^\downarrow - E_o^\downarrow} \quad (1)$$

$$\Delta E^{\uparrow\downarrow} = \Delta E^{\uparrow\downarrow}(x) - \Delta E^{\uparrow\downarrow}(z)$$

$$= -\xi^2 \sum_{o^\uparrow, u^\downarrow} \frac{\left|\langle o^\uparrow |L_z| u^\downarrow \rangle\right|^2 - \left|\langle o^\uparrow |L_x| u^\downarrow \rangle\right|^2}{E_u^\downarrow - E_o^\uparrow} \quad (2)$$

Where, ξ is the SOC constant, u and o are the unoccupied states and occupied states, respectively. The superscript of u and o, ↑ and ↓, represent the corresponding spin-majority state and spin-minority states, respectively. Eq. (1) is the spin-conservation term; Eq. (2) is the spin-flip term. According to the Eq. (1) and (2), the MAE is determined by the spin-orbital matrix element differences, their energy differences as well as the states density. The nonzero *Lz* and *Lx* matrix elements are $\langle d_{xz}|L_z|d_{yz}\rangle=1$, $\langle d_{xy}|L_z|d_{x^2-y^2}\rangle=2$, $\langle d_{z^2}|L_x|d_{xz},d_{yz}\rangle=\sqrt{3}$, $\langle d_{xy}|L_x|d_{xz},d_{yz}\rangle=1$ and $\langle d_{x^2-y^2}|L_x|d_{xz},d_{yz}\rangle=1$.[25] In terms of Eq. (1) and (2), for $d_{xy}/d_{x2-y2}$ and $d_{xz}/d_{yz}$ mixed orbitals, the spin-conservation term contributes positive MAE, while the spin-flip term contributes the negative MAE. Differently, for the others mixed orbitals, their spin-conservation term and spin-flip term contribute opposite MAE relative to the $d_{xy}/d_{x2-y2}$ and $d_{xz}/d_{yz}$ orbitals.

It is noted that, for $d_{xy}/d_{x2-y2}$ and $d_{xz}/d_{yz}$ orbitals, except for Co2 atom in $In_2Se_3$-CoCo and Os atom in $In_2Se_3$-CoOs, all the other atoms of the dimers in the four systems show the same changing trend of MAE with FE polarization switching. When the FE polarization switches from P↑ to P↓, the values of MAE from $d_{xy}/d_{x2-y2}$ and $d_{xz}/d_{yz}$ orbitals increase. The main reason for this trend can be attributed the decrease of the MAE contributed from the spin-flip term of these orbitals. One of the key parameters to determine the value of the spin-flip term for $d_{xy}/d_{x2-y2}$ and $d_{xz}/d_{yz}$ is the exchange splitting, which is the energy difference between spin-minority state and spin-majority state.[26] Large exchange splitting will reduce the absolute value of the negative MAE from the spin-flip term for $d_{xy}/d_{x2-y2}$ and $d_{xz}/d_{yz}$ orbitals, and hence increase the total MAE. When the absolute value of positive part of MAE becomes larger than that of the negative part of MAE, the MA easy axis turns into out-of-plane direction. To confirm the effects of exchange splitting on the MAE of $d_{xy}/d_{x2-y2}$ and $d_{xz}/d_{yz}$ orbitals, we calculated the exchange splitting from the orbital-resolved DOS. Here, we determine the exchange splitting by the difference between the energy band centers of spin-minority state and spin-majority state. The energy band centers were calculated with the formula

$$d-center = \frac{\int E\rho dE}{\int \rho dE} \quad (3)$$

where, the $\rho$ is the DOS of the d orbitals at E energy. The calculated results are shown in Table 2. We can see that, except for Co2 in $In_2Se_3$-CoCo and Os in $In_2Se_3$-CoOs, the exchange splittings of $d_{xy}/d_{x2-y2}$ and $d_{xz}/d_{yz}$ orbitals for the other atoms indeed increase with switching FE polarization from P↑ to P↓. Thus, for the four systems, one of main origins for the increase of MAE as FE polarization changes from P↑to P↓ is the enlargement of the exchange splitting of $d_{xy}/d_{x2-y2}$ and $d_{xz}/d_{yz}$ orbitals. It is worth mentioning that the calculated exchange splittings are also in agreement with the calculated values of magnetic moments. As shown in Figures 2(b)-(e), except for Co2 in $In_2Se_3$-CoCo and Os in $In_2Se_3$-CoOs, the magnetic moments from $d_{xy}/d_{x2-y2}$ and $d_{xz}/d_{yz}$ orbitals of the other atoms almost all increase as FE polarization changes from P↑ to P↓.

**Table 2** Exchange splittings of $d_{xy}/d_{x2-y2}$ and $d_{xz}/d_{yz}$ orbitals in $In_2Se_3$-CoCo, $In_2Se_3$-OsOs, $In_2Se_3$-CoOs and $In_2Se_3$-OsCo with FE polarization P↑ and P↓.

|  |  |  | $In_2Se_3$-CoCo | | $In_2Se_3$-OsOs | | $In_2Se_3$-CoOs | | $In_2Se_3$-OsCo | |
| --- | --- | --- | --- | --- | --- | --- | --- | --- | --- | --- |
|  |  |  | Co1 | Co2 | Os1 | Os2 | Co | Os | Os | Co |
| $\Delta E_{ex}$ | P↑ | $d_{xy}/d_{x2-y2}$ | 1.51 | 2.01 | 0 | 0.53 | 1.05 | 2.30 | 0.04 | 1.84 |
|  |  | $d_{xz}/d_{yz}$ | 1.73 | 2.33 | 0.01 | 0.32 | 1.31 | 1.87 | 0.19 | 1.72 |
|  | P↓ | $d_{xy}/d_{x2-y2}$ | 1.59 | 1.97 | 0.04 | 1.05 | 1.16 | 2.16 | 0.08 | 1.85 |
|  |  | $d_{xz}/d_{yz}$ | 1.82 | 2.16 | 0.02 | 0.63 | 1.34 | 1.78 | 0.27 | 1.74 |

For the Co2 atom in $In_2Se_3$-CoCo, although the MAE of $d_{xy}/d_{x2-y2}$ and $d_{xz}/d_{yz}$ orbitals decrease with FE polarization changing from P↑to P↓, the MAE from the $d_{z2}/d_{yz}$ and $d_{z2}/d_{xz}$ orbitals increase, resulting in the enhancement of the total MAE. From the Figure 3(a), it can be seen that, when FE polarization changes from P↑to P↓, the $d_{z2}$ spin-minority state of Co2 atom around Fermi level moves upward and the occupation state below Fermi level decrease, resulting in the reduction of absolute value of MAE from the spin-conservation term of $d_{z2}/d_{yz}$ and $d_{z2}/d_{xz}$. The upward shift of spin-minority states of $d_{z2}$ orbital for Co2 atom can also be confirmed by the results of the calculated magnetic moments. As shown in Figure 2(b), the magnetic moments of $d_{z2}$ orbital increase with FE polarization switching to P↓. This can only be caused by the upward shift of spin-minority state of $d_{z2}$ orbital, since the spin-majority state all distribute below the Fermi level for both P↑ to P↓, as shown in Figure 3(a). For the Os atom in $In_2Se_3$-CoOs, the increase of MAE with FE polarization changing from P↑ to P↓ is also related with the $d_{z2}$ orbital. Whereas, the difference is that the increase of MAE mainly originates from the changes of spin-flip term of $d_{z2}/d_{xz}$ orbitals. As shown in Figure 3(c), in comparison with the P↑, the spin-majority state of $d_{z2}$ orbital with P↓ move upward, leading to the occupation in the vicinity of Fermi level increase. As a result, the positive MAE from spin-flip term of $d_{z2}/d_{xz}$ orbital increases, and eventually induces the enhancement of the perpendicular MA in $In_2Se_3$-CoOs.

## 4. CONCLUSION

In summary, by first-principle calculations, we have studied the magnetism of dimers, Co-Os, Co-Co and OsOs, adsorbed on the 2D ferroelectric $In_2Se_3$. We find that the Co-Os dimer in $In_2Se_3$-CoOs configuration has high MAE of 40 meV. Especially, the MAE of Os atom is up to ~60 meV due to its special coordination and large SOC constant. Our calculated results also reveal that, via switching the FE polarization of $In_2Se_3$, the MA and magnetic moments of the dimers on $In_2Se_3$ can be tuned. Using second-order perturbation theory, we find that the tunable of MA and magnetic moments is related to the exchange splitting of $d_{xy}/d_{x2-y2}$ and $d_{xz}/d_{yz}$ orbitals as well as the occupation of $d_{z2}$ orbital around Fermi level. This work provides important theoretical insights for the MA of transition metal dimers on 2D ferroelectric $In_2Se_3$, and may further stimulate theoretical and experimental exploration in related fields.

**Notes**

The authors declare no competing financial interest.

**Acknowledgements**

This work was supported by the Open Project of Key Laboratory for Magnetism and Magnetic Materials of the Ministry of Education, Lanzhou University (Grant No. LZUMMM2021006), the Open Project of National Laboratory of Solid State Microstructures, Nanjing University (Grant No. M33010) and the School Scientific Research Project of Hangzhou Dianzi University (Grant Nos. KYS045619084, KYS045619085).